\documentclass[11pt]{article}
\usepackage{amsmath,amssymb,amsthm,amsxtra,overpic,bbm,bm,epsfig,ulem,color,multirow}
\usepackage{braket}
\usepackage{hyperref}

\begin{document}

\begin{center}
{\Large Linear seesaw leptogenesis before/after electroweak symmetry breaking}
\end{center}

\vspace{0.05cm}

\begin{center}
{\bf Yan Shao, \bf Zhen-hua Zhao\footnote{Corresponding author: zhaozhenhua@lnnu.edu.cn}} \\
{ $^1$ Department of Physics, Liaoning Normal University, Dalian 116029, China \\
$^2$ Center for Theoretical and Experimental High Energy Physics, \\ Liaoning Normal University, Dalian 116029, China }
\end{center}

\vspace{0.2cm}

\begin{abstract}
The linear seesaw (LSS) model provides a natural framework for generating small neutrino masses at low energy scales, thereby offering promising testability prospects. However, in generic LSS models, the exact mass degeneracy (before the electroweak symmetry breaking) between the two sterile neutrinos that form a Dirac pair precludes the generation of CP asymmetries from their interplay, posing a significant challenge to explaining the observed baryon (or lepton) asymmetry of the Universe via the leptogenesis mechanism. In this work, we explore two well-motivated approaches to generate a suitable mass splitting for the two sterile neutrinos that form a Dirac pair, and consequently naturally realize a resonantly enhanced generation of baryon (and lepton) asymmetry. First, we demonstrate that the renormalization group evolution effects can naturally induce the desired mass splitting for the sterile neutrinos, resulting in a successful generation of the observed baryon asymmetry of the Universe. Second, motivated by the recent result from the EMPRESS collaboration that indicates the possible existence of a large lepton asymmetry of the Universe, we explore the possibility that a large lepton asymmetry might naturally follow from the electroweak symmetry breaking which automatically induces the desired mass splitting for the sterile neutrinos.

\end{abstract}

\newpage

\section{Introduction}

The origin of the baryon-antibaryon asymmetry of the Universe, which is quantified as \cite{Planck:2018vyg}
\begin{eqnarray}
Y^{}_{\rm B} \equiv \frac{n^{}_{\rm B}-n^{}_{\rm \bar B}}{s} \simeq (8.68 \pm 0.06) \times 10^{-11}  \;,
\label{1.1}
\end{eqnarray}
where $n^{}_{\rm B}$ ($n^{}_{\rm \bar B}$) denotes the baryon (antibaryon) number density and $s$ the entropy density, remains one of the major unresolved puzzles in particle cosmology. Among various theoretical explanations, the leptogenesis mechanism \cite{leptogenesis}-\cite{Lreview4} stands out as an elegant and compelling scenario. In this mechanism, a lepton asymmetry $Y^{}_{\rm L} \equiv (n^{}_{\rm L} -n^{}_{\rm \bar L})/s$ (defined in the same way as $Y^{}_{\rm B}$) is first generated from the CP-violating and out-of-equilibrium decays of right-handed neutrinos (RHNs) $N^{}_I$ (for $I=1, 2, ...$), which are typically introduced in the type-I seesaw extension of the Standard Model \cite{seesaw1}-\cite{seesaw5}. The lepton asymmetry is subsequently partially converted to the observed baryon asymmetry via the sphaleron processes: $Y^{}_{\rm B} \simeq -1/3 Y^{}_{\rm L}$. It should be noted that the sphaleron processes only maintain equilibrium within the temperature range of [132, $10^{12}$] GeV \cite{sphaleron}.

The type-I seesaw model thus provides a unified explanation for both the smallness of neutrino masses and the baryon asymmetry of the Universe. However, it faces a significant challenge in terms of experimental testability: the newly introduced particle states (i.e., RHNs) are too heavy to be directly accessed by foreseeable experiments: on the one hand, in order to naturally generate eV-scale neutrino masses, the RHN masses need to be around ${\cal O}(10^{13})$ GeV. On the other hand, in order to generate the observed baryon asymmetry, the RHN masses usually need to be above ${\cal O}(10^9)$ GeV \cite{DI}, unless they are nearly degenerate (in which case the resonant leptogenesis scenario \cite{resonant1, resonant2} can be realized and successfully reproduce the observed baryon asymmetry at low energy scales). This inherent tension between naturalness and testability motivates the exploration of alternative mechanisms that are capable of generating the neutrino masses and observed baryon asymmetry at low energy scales.

The linear seesaw (LSS) model \cite{LSS1}-\cite{LSS3} provides an alternative framework, which  introduces extra singlet fermions $S_I$ (for $I=1,2,\ldots$) in addition to the RHNs. The relevant Lagrangian terms contributing to the neutrino mass generation are given by
\begin{eqnarray}
-\mathcal{L}^{}_{\rm mass} \supset (Y^{}_{\nu})^{}_{\alpha I} \, \overline{L^{}_\alpha} \, \widetilde{H} \, N^{}_I + (Y^{}_{\rm LS})^{}_{\alpha I} \, \overline{L^{}_\alpha} \, \widetilde{H} \, S^{c}_I + (M^{}_{\rm RS})^{}_{IJ} \, \overline{N^c_{I}} \, S^{c}_{J}  + {\rm h.c.}
\label{1.2}
\end{eqnarray}
where $L^{}_\alpha$ (for $\alpha = e, \mu, \tau$) and $H$ (for $\widetilde{H} = {\rm i \sigma^{}_2} H^*$ with $\sigma^{}_2$ being the second Pauli matrix) respectively denote the lepton and Higgs doublets, and the superscript $c$ denotes the charge conjugated fields. And $Y^{}_\nu$, $Y^{}_{\rm LS}$ and $M^{}_{\rm RS}$ are Yukawa and mass matrices in the flavor space. After the Higgs field acquires the vacuum expectation value $v=174$ GeV (i.e., the electoweak symmetry breaking), the Yukawa couplings will lead to some mass terms connecting the left-handed neutrinos $\nu^{}_\alpha$ to both RHNs and $S_I$: $(M^{}_{\rm D})^{}_{\alpha I} = (Y^{}_{\nu})^{}_{\alpha I} v$ and  $(M^{}_{\rm LS})^{}_{\alpha I} = (Y^{}_{\rm LS})^{}_{\alpha I} v$. In the $(\nu_{\alpha}^c, N^{}_{I}, S_{I}^c)$ basis, the complete neutrino mass matrix takes the form
\begin{eqnarray}
M^{}_{\nu NS} = \begin{pmatrix}
0 & M^{}_{\rm D} & M^{}_{\rm LS} \\
M^{T}_{\rm D} & 0 & M^{}_{\rm RS} \\
M^{T}_{\rm LS} & M^{T}_{\rm RS} & 0 \\
\end{pmatrix} \;.
\label{1.3}
\end{eqnarray}
In this model the lepton number is violated by the Yukawa couplings in $Y^{}_{\rm LS}$, so the latter can be naturally small (i.e., the lepton number symmetry would get restored in the limit of $Y^{}_{\rm LS} \rightarrow 0$).
Under the natural condition of $M^{}_{\rm LS} \ll M^{}_{\rm D} \ll M^{}_{\rm RS}$, block diagonalization of Eq.~(\ref{1.3}) yields the light neutrino mass matrix as follows:
\begin{equation}
M^{}_{\nu} \simeq -[ M^{}_{\rm D} (M^{}_{\rm LS} M^{-1}_{\rm RS})^{T} + (M^{}_{\rm LS} M^{-1}_{\rm RS}) M^{T}_{\rm D}] \;.
\label{1.4}
\end{equation}
This shows that the smallness of neutrino masses can naturally follow from small values of $M^{}_{\rm LS}$, rendering superheavy sterile neutrinos no longer necessary and a direct detection of them possible \cite{FMN}. Without loss of generality, we will work in the basis that $M^{}_{\rm RS}$ is diagonal as $M^{}_{\rm RS} = {\rm diag}(M^{}_1, M^{}_2, M^{}_3)$.

On the other hand, the RHN and $S^{}_I$ singlets (will be collectively referred to as sterile neutrinos) will form Dirac pairs. A Dirac pair can be thought of as two exactly degenerate Majorana sterile neutrinos with opposite CP phases: diagonalization of the sterile neutrino mass matrix (i.e., the bottom-right sub-matrix of Eq.~(\ref{1.3}), denoted as $M^{}_{NS}$) yields the mass eigenvalues $-M^{}_1, -M^{}_2, -M^{}_3, M^{}_1, M^{}_2$ and $M^{}_3$, which means that $N^{}_1$ and $S^{}_1$ (and similarly for $N^{}_2$ and $S^{}_2$, and $N^{}_3$ and $S^{}_3$) form a Dirac pair.  Unfortunately, the exact mass degeneracy between the two sterile neutrinos that form a Dirac pair precludes the generation of CP asymmetries from their interplay, posing a significant challenge to explaining the observed baryon (or lepton) asymmetry of the Universe via the leptogenesis mechanism. But just as a coin has two sides, such a feature of the LSS model can serve as a unique starting point for the realization of resonant leptogenesis scenarios \cite{resonant1, resonant2}: if the sterile neutrino masses receive certain corrections so that their degeneracy is lifted to an appropriate extent \footnote{After a Dirac pair of sterile neutrinos receive a small mass splitting, they will be called a pseudo-Dirac pair \cite{pseudo1}-\cite{pseudo4}.}, then the resonant leptogenesis scenario will be naturally realized, thereby enabling successful leptogenesis at low energy scales.

In this work, we explore two well-motivated approaches to generate the desired mass splitting for the two sterile neutrinos that form a Dirac pair: the renormalization group evolution (RGE) effect and the electroweak symmetry breaking (EWSB) (see sections~3 and 4 for details). These two approaches have the following three common merits: 1) the mass splittings for the sterile neutrinos such generated are typically tiny, which are just suitable for triggering resonant leptogenesis. 2) These two effects are spontaneous and unaovidable. 3) These two effects are minimal in the sense that they do not need to introduce additional parameters. But there exists a crucial difference between the two approaches: when the latter approach begins to take effect (at $T \simeq 156$ GeV) the sphaleron processes will freeze out almost at the same time (at $T \simeq 132$ GeV \cite{sphaleron}). This means that the lepton asymmetry generated from the latter approach probably will not get converted into a baryon asymmetry of the same order of magnitude. We will just make use of such a feature of the LSS model to give a tentative explanation for the possible existence of a large lepton asymmetry as indicated by the recent result from the EMPRESS collaboration (see section~4.2), without contradicting the observed baryon asymmetry.

The remaining parts of this paper are organized as follows. In the next section, we recapitulate some basic formulas of leptogenesis as a basis of our study. In section~3, we demonstrate that the RGE effects can naturally induce the desired mass splitting for the sterile neutrinos, resulting in a successful generation of the observed baryon asymmetry of the Universe. In section~4, we show that the observed baryon asymmetry of the Universe can also be naturally generated through the EWSB, which induces the desired mass splitting for the sterile neutrinos shortly before the sphaleron freeze-out. Furthermore, motivated by the recent result from the EMPRESS collaboration that indicates the possible existence of a large lepton asymmetry of the Universe, we explore the possibility that the EWSB-induced mass splitting may generate a large lepton asymmetry after the sphaleron freeze-out. For these scenarios, we will also investigate the consequences of the model with respect to the charged lepton flavour violation processes. Finally, the summary of our main results will be given in section~5.

\section{Some basic formulas of leptogenesis}

In this section, we recapitulate some basic formulas of leptogenesis as a basis of our study.

For temperatures of the Universe below $\sim 10^{9}$ GeV, the three lepton flavors (for $\alpha = e, \mu, \tau$) are effectively distinguishable. Therefore, in the low-scale (TeV or so) leptogenesis regime relevant to our study, it is necessary to treat the lepton asymmetries stored in each flavor separately \cite{flavor1, flavor2}. In the meantime, the contributions of the nearly degenerate sterile neutrinos to the final baryon asymmetry are on equal footing and should be taken into consideration altogether \cite{resonant1, resonant2}. For these reasons, the final baryon asymmetry can be expressed as
\begin{eqnarray}
Y^{}_{\rm B}  = c Y^{}_{\rm L} = c r \sum^{}_\alpha \varepsilon^{}_{\alpha} \kappa^{}_\alpha \;,
\label{2.1}
\end{eqnarray}
where $c \simeq -1/3$ is the conversion efficiency from the lepton asymmetry to the baryon asymmetry via the sphaleron processes, $r \simeq 4 \times 10^{-3}$ measures the ratio of the equilibrium number density of sterile neutrinos to the entropy density, and $\varepsilon^{}_\alpha$ is a sum over the sterile neutrinos (i.e., $\varepsilon^{}_\alpha= \sum^{}_{I}\varepsilon^{}_{I \alpha}$) of the CP asymmetries for their decays. To compute the flavor-specific CP asymmetries $\varepsilon^{}_{I\alpha}$, we work in the mass basis of sterile neutrinos, where the effective Yukawa coupling matrix is defined as $h^{}_{\nu}= (Y^{\prime}_{\nu}, Y^{\prime}_{\rm LS})$, with $Y^{\prime}_{\nu}$ and $Y^{\prime}_{\rm LS}$ denoting the $3 \times 3$ Yukawa matrices in the mass basis of sterile neutrinos (see the discussions around Eq.~(\ref{3.6})). In the resonant leptogenesis regime, the CP asymmetries for the decays of the $I$-th sterile neutrino are given by \cite{resonant1, resonant2}
\begin{eqnarray}
&& \varepsilon^{}_{I\alpha} =  \sum^{}_{J \neq I} \varepsilon^{}_{IJ\alpha}
=  \sum^{}_{J \neq I} \frac{1}{8\pi (h^\dagger_{\rm \nu} h^{}_{\rm \nu})^{}_{II}} {\rm Im}\left[ (h^*_{\rm \nu})^{}_{\alpha I} (h^{}_{\rm \nu})^{}_{\alpha J}
( (h^\dagger_{\rm \nu} h^{}_{\rm \nu})^{}_{IJ} + \frac{M_I}{M_J} (h^\dagger_{\rm \nu} h^{}_{\rm \nu})^{}_{JI} ) \right]  \cdot f^{}_{IJ} \;, \nonumber \\
&& {\rm with } \hspace{1cm} f^{}_{IJ} \equiv \frac{M^{}_I M^{}_J  \Delta M^2_{IJ}}{(\Delta M^2_{IJ})^2 + M^2_I \Gamma^2_J} \;,
\label{2.2}
\end{eqnarray}
where $\Delta M^2_{IJ} \equiv M^2_I - M^2_J$, and $\Gamma^{}_J= \sum^{}_{\alpha} \Gamma^{}_{J\alpha} = \sum^{}_{\alpha} (h^\dagger_{\rm \nu} h^{}_{\rm \nu})^{}_{J\alpha} M^{}_J/8\pi$ is the decay rate of the $J$-th sterile neutrino.
Last but not least, $\kappa^{}_\alpha$ are the flavor-specific efficiency factors that take account of the washout effects due to the inverse decays of sterile neutrinos and various lepton-number-violating scattering processes.

In general, the strengths of washout effects are governed by the following flavor-specific washout parameters
\begin{eqnarray}
K^{}_{\alpha}= \sum^{}_{I} K_{I \alpha} =  \sum^{}_{I} \frac{\Gamma^{}_{I \alpha}}{H(T=M^{}_{I})} \;,
\label{2.3}
\end{eqnarray}
where $\Gamma^{}_{I \alpha}$ has been defined below Eq.~(\ref{2.2}) and $H(T) =1.66 \sqrt{g_*}\, T^2/M^{}_{\rm Pl}$ is the Hubble rate with $T$ being the temperature of the thermal bath, $g^{}_*$ the number of relativistic degrees of freedom and $M^{}_{\rm Pl} = 1.22\times 10^{19}$ GeV the Planck mass.
However, in seesaw models with approximate lepton number conservation (such as the LSS model considered here), the usually-ignored interference term involving different sterile neutrinos may become essential \cite{Yv}. This motivates the definition of an effective washout parameter \cite{Yv}
\begin{eqnarray}
K^{\rm eff}_{I \alpha} \equiv K^{}_{I \alpha} \cdot \frac{\delta^{2}_{}}{1+\sqrt{a_{}}\delta_{}+\delta^{2}_{}} \;,
\label{2.4}
\end{eqnarray}
where $\delta \equiv |\Delta M|/\Gamma$ with $\Delta M$ being the mass splitting within each pair of pseudo-Dirac sterile neutrinos, and $a= (\Gamma^{}_{} /M^{}_{})^2$. This additional factor arises from the following observations. On the one hand, in models with approximately conserved lepton number such as the LSS model considered here, washout effects (which inherently violate the lepton number) vanish in the limit where the lepton-number-violating parameters (specifically, $Y^{}_{\rm LS}$) approach zero. On the other hand, the mass splitting within each pseudo-Dirac sterile neutrino pair is also controlled by these same lepton-number-violating parameters (again, $Y^{}_{\rm LS}$, as discussed in sections~3 and 4). Consequently, the effective washout parameter decreases to zero as the mass splitting $\Delta M$ tends to zero, in accordance with Eq.~(\ref{2.4}) \cite{Yv}.

The overall impact of washout processes is determined by the total effective washout parameter for each flavor, i.e., $K^{\rm eff}_{\alpha}= \sum^{}_{I} K^{\rm eff}_{I \alpha}$. Depending on its magnitude, the efficiency factor $\kappa_\alpha$ exhibits two qualitatively distinct behaviors, corresponding to the strong and weak washout regimes. In the strong washout regime where $K^{\rm eff}_{\alpha} \geqslant 1$, the generated lepton asymmetry is significantly suppressed by inverse processes. The efficiency factor can be approximated by
\begin{eqnarray}
\kappa^{}_{\alpha} \approx \frac{2}{z^{}_{B} K^{\rm eff}_{\alpha}}\left[1-e^{-\frac{1}{2}z^{}_{B} K^{\rm eff}_{\alpha}}\right] \;,
\label{2.5}
\end{eqnarray}
where $z^{}_{B}$ is well approximated by \cite{zB}
\begin{eqnarray}
z^{}_{B} \approx  1+\frac{1}{2}\ln\left[1+\frac{\pi {K^{\rm eff}_{\alpha}}^2}{1024}\left(\ln\frac{3125\pi {K^{\rm eff}_{\alpha}}^2}{1024}\right)^{5}\right] \;.
\label{2.6}
\end{eqnarray}
In the weak washout regime where $K^{\rm eff}_{\alpha} < 1 $, the generated lepton asymmetry is less affected by inverse processes. And the final efficiency depends on whether $K^{}_{\alpha}$ is larger or smaller than 1, leading to the following two distinct cases \cite{ADEFHV}
\begin{eqnarray}
\kappa_{\alpha} \approx
\left\{
\begin{array}{ll}
\frac{1}{4} K^{\rm eff}_{\alpha} \left(\frac{3\pi}{2} - z^{}_{B}\right) \hspace{0.85cm} (K^{}_{\alpha} > 1) \;; \\
\frac{9\pi^{2}}{64} K^{\rm eff}_{\alpha} K^{}_{\alpha} \hspace{1.8cm} (K^{}_{\alpha} < 1) \;.
\end{array}
\right.
\label{2.7}
\end{eqnarray}

\section{Renormalization group evolution assisted leptogenesis}

In this section, we study the scenario that the desired mass splitting for the sterile neutrinos that form a Dirac pair is generated from the RGE effect.

\subsection{Consequence for leptogenesis}

In the SM framework, the one-loop RGE equations of the neutrino sector parameters are given by \cite{Brdar:2015jwo}
\begin{eqnarray}
&& 16\pi^2  \frac{{\rm d}Y^{}_{\nu}}{{\rm d}t}  =\left[ \alpha^{}_\nu -\frac{3}{2} Y^{}_l Y^\dagger_l + \frac{3}{2} (Y^{}_{\nu} Y^\dagger_{\nu}+Y^{}_{\rm LS} Y^\dagger_{\rm LS}) \right]Y^{}_{\nu} \; ,
 \nonumber \\
&& 16\pi^2  \frac{{\rm d}Y^{}_{\rm LS}}{{\rm d}t}  =\left[ \alpha^{}_\nu -\frac{3}{2} Y^{}_l Y^\dagger_l + \frac{3}{2} (Y^{}_{\nu} Y^\dagger_{\nu}+Y^{}_{\rm LS} Y^\dagger_{\rm LS}) \right]Y^{}_{\rm LS} \; ,
 \nonumber \\
&& 16\pi^2  \frac{{\rm d}M^{}_{\rm R}}{{\rm d}t}  =(Y^\dagger_{\nu} Y^{}_{\nu}) M^{}_{\rm R} +   M^{}_{\rm R} (Y^\dagger_{\rm LS} Y^{}_{\rm LS} )^T\; , \nonumber \\
&& 16\pi^2  \frac{{\rm d}\mu^{}_n}{{\rm d}t}  =(Y^\dagger_{\nu} Y^{}_{\rm LS}) M^{{\rm T}}_{\rm R} +   M^{}_{\rm R} (Y^\dagger_{\nu} Y^{}_{\rm LS} )^T\; , \nonumber \\
&& 16\pi^2  \frac{{\rm d}\mu^{}_s}{{\rm d}t}  =(Y^\dagger_{\rm LS} Y^{}_{\nu}) M^{}_{\rm R} +   M^{{\rm T}}_{\rm R} (Y^\dagger_{\rm LS} Y^{}_{\nu} )^T\; ,
\label{3.1}
\end{eqnarray}
where $t = \ln \left(\mu/ \Lambda^{}_{}\right)$ has been defined with $\mu$ being the renormalization scale, and $\alpha_\nu $ is defined as
\begin{eqnarray}
\alpha^{}_\nu  =  {\rm tr} \left(3 Y^{}_u Y^\dagger_u + 3 Y^{}_d Y^\dagger_d + Y^{}_l Y^\dagger_l +  Y^{}_{\nu} Y^\dagger_{\nu}+Y^{}_{\rm LS} Y^\dagger_{\rm LS} \right)-\frac{9}{20} g^2_1 - \frac{9}{4} g^2_2  \;.
\label{3.2}
\end{eqnarray}
Here $g^{}_1$ and $g^{}_2$ are respectively  the $\rm U\left(1\right)^{}_{\rm Y}$ and $\rm SU\left(2\right)^{}_{\rm L}$ gauge coupling constants, and $Y^{}_{u,d,l}$ are respectively the Yukawa matrices for up-type quarks, down-type quarks and charged leptons.
A particularly important consequence of RGE in the LSS model is the dynamical generation of mass terms $\mu_n$ and $\mu_s$ (see the last two equations of Eq.~(\ref{3.1})), which originally vanish at the tree level. They play the crucial role in breaking the mass degeneracy of sterile neutrinos that form a Dirac pair, thereby enabling the resonant enhancement necessary for a successful leptogenesis. After including these terms, the sterile neutrino mass matrix is modified into the following form:
\begin{eqnarray}
M_{NS}^{\prime} = \begin{pmatrix}
 \mu^{}_n  & M^{}_{\rm RS} \\
 {M^{}_{\rm RS}}^{\rm T}& \mu^{}_s \end{pmatrix} \;.
\label{3.3}
\end{eqnarray}

To facilitate the leptogenesis calculation, it is convenient to transform to the sterile neutrino mass basis, in which $ M^{\prime}_{NS}$ becomes real, positive and diagonal. For this purpose, one can diagonalize it with the help of a unitary transformation $V$ as follows
\begin{eqnarray}
V^{T}_{} M^{\prime}_{NS} V^{}_{} \approx
\begin{pmatrix}
V^{T}_{1}\left[-M^{}_{\rm RS}+\frac{\mu^{}_{\rm s}+\mu^{}_{\rm n}}{2} \right] V^{}_{1} &  0  \\
0	&  V^{T}_{2}\left[M^{}_{\rm RS}+\frac{\mu^{}_{\rm s}+\mu^{}_{\rm n}}{2}  \right] V^{}_{2}
\end{pmatrix} \;,
\label{3.4}
\end{eqnarray}
with
\begin{eqnarray}
V^{}_{}\approx\frac{1}{\sqrt{2}}
\begin{pmatrix}
\mathbf{1}+\frac{\left(\mu^{}_{\rm s}-\mu^{}_{\rm n}\right) {M^{-1}_{\rm RS}}^{}}{4}	& \mathbf{1}-\frac{\left(\mu^{}_{\rm s}-\mu^{}_{\rm n}\right) {M^{-1}_{\rm RS}}^{}}{4} \\
-\mathbf{1}+\frac{\left(\mu^{}_{\rm s}-\mu^{}_{\rm n}\right) {M^{-1}_{\rm RS}}^{}}{4} & \mathbf{1}+\frac{\left(\mu^{}_{\rm s}-\mu^{}_{\rm n}\right) {M^{-1}_{\rm RS}}^{}}{4}
\end{pmatrix}
\begin{pmatrix}
V^{}_{1} & 0 \\
0  & V^{}_{2}
\end{pmatrix}  \;,
\label{3.5}
\end{eqnarray}
where $V^{}_{1}$ and $V^{}_{2}$ are the unitary matrices for diagonalizing $-M^{}_{\rm RS}+\frac{\mu^{}_{\rm s}+\mu^{}_{\rm n}}{2}$ and $M^{}_{\rm RS}+\frac{\mu^{}_{\rm s}+\mu^{}_{\rm n}}{2}$, respectively.
After this unitary transformation, the complete neutrino mass matrix becomes
\begin{eqnarray}
M_{\nu NS}^{\prime} \simeq \begin{pmatrix}
0 & v Y^{\prime}_{\nu} & v Y^{\prime}_{\rm LS} \\
v \left(Y^{\prime}_{\nu}\right)^{T} & V^{T}_{1}\left[-M^{}_{\rm RS}+\frac{\mu^{}_{\rm s}+\mu^{}_{\rm n}}{2} \right] V^{}_{1} & 0 \\
v \left(Y^{\prime}_{\rm LS}\right)^{T} & 0 & V^{T}_{2}\left[M^{}_{\rm RS}+\frac{\mu^{}_{\rm s}+\mu^{}_{\rm n}}{2}   \right] V^{}_{2} \end{pmatrix} \;,
\label{3.6}
\end{eqnarray}
where $Y^{\prime}_{\nu}$ and $Y^{\prime}_{\rm LS}$ correspond to the rotated $3\times 3$ Yukawa matrices (mentioned above Eq.~(\ref{2.2})) that couple the sterile neutrinos to the SM leptons, and they can be analytically expressed as follows \cite{Yv}
\begin{eqnarray}
&& Y^{\prime}_{\nu}  \simeq  \frac{1}{\sqrt{2}\,v}\left[
M^{}_{\rm D}-M^{}_{\rm LS}+\frac{\left( M^{}_{\rm D}+M^{}_{\rm LS}\right)\left(\mu^{}_{\rm s}-\mu^{}_{\rm n}\right) {M^{-1}_{\rm RS}}}{4}  \right] V^{}_{1} \;,
\nonumber\\
&& Y^{\prime}_{\rm LS}  \simeq  \frac{1}{\sqrt{2}\,v}\left[
M^{}_{\rm D}+M^{}_{\rm LS}-\frac{\left( M^{}_{\rm D}-M^{}_{\rm LS}\right)\left(\mu^{}_{\rm s}-\mu^{}_{\rm n}\right) {M^{-1}_{\rm RS}}}{4}  \right] V^{}_{2} \;.
\label{3.7}
\end{eqnarray}

To ensure consistency of our parameter choices with the neutrino oscillation data (see the global-fit results for the neutrino-oscillation experiments in Refs.~\cite{global1, global2}), we employ the following modified Casas-Ibarra parametrization~\cite{CI} as applicable to the LSS model \cite{Forero:2011pc} \footnote{For a full parametrization of the neutrino flavor mixing matrix for the LSS model in terms of 36 rotation angles and 36 CP-violating phases, see Ref.~\cite{xing}.} to reconstruct the Dirac neutrino mass matrix $M^{}_{\rm D}$:
\begin{equation}
M^{}_{\rm D} = - U\, D_{\nu}^{1/2}\, R\, D_{\nu}^{1/2}\, U^{T}\, \left(M^{T}_{\rm LS}\right)^{-1} M^{T}_{\rm RS} \; ,
\label{3.10}
\end{equation}
where $D^{}_\nu = {\rm diag}(m^{}_1, m^{}_2, m^{}_3)$ with $m^{}_i$ being three light neutrino masses, and $U$ is the PMNS neutrino mixing matrix \cite{xingPR}. The matrix $R$ takes a form as
\begin{equation}
R = \begin{pmatrix}
\frac{1}{2} & a & b \\
-a & \frac{1}{2} & c \\
-b & -c & \frac{1}{2} \end{pmatrix} \, ,
\label{3.11}
\end{equation}
where $a$, $b$ and $c$ are generally complex parameters.

\begin{figure*}
\centering
\includegraphics[width=6.5in]{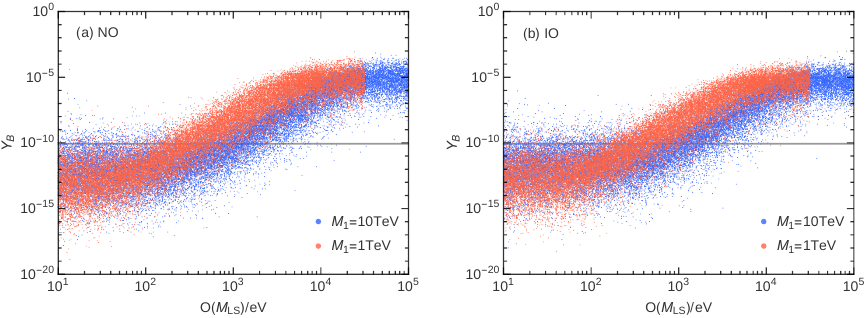}
\caption{ For the scenario studied in section~3.1, in the NO (a) and IO (b) cases, the allowed values of $Y^{}_{\rm B}$ as functions of ${\cal O}(M^{}_{\rm LS})$ for some benchmark values of $M^{}_1$. The horizontal line stands for the observed value of $Y^{}_{\rm B}$. }
\label{fig1}
\end{figure*}

Now, we are ready to perform the numerical calculations. In this paper, we consider the scenario that the three sterile neutrino mass parameters $M^{}_1, M^{}_2$ and $M^{}_3$ are hierarchical, and thus only need to consider the contributions of the lightest pseudo-Dirac pair of sterile neutrinos (formed by $N^{}_1$ and $S^{}_1$) to leptogenesis.
In Figure~\ref{fig1}(a) and (b) (for the normal ordering (NO) and inverted ordering (IO) cases of light neutrino masses, respectively) we have shown the allowed values of $Y^{}_{\rm B}$ as functions of ${\cal O}(M^{}_{\rm LS})$ (the order of magnitude of $M^{}_{\rm LS}$) for the benchmark values of $M^{}_1 = 1$ and 10 TeV. These results are obtained for the following parameter settings: for the neutrino mass-squared differences and mixing angles, we adopt the values from global-fit analyses \cite{global1, global2}. The mass of the lightest neutrino (either $m^{}_1$ in the NO case or $m^{}_3$ in the IO case) is varied over the range 0.001---0.1 eV. The CP-violating phases in the neutrino mixing matrix are allowed to vary freely between 0 and $2\pi$. For the entries of the matrix $M^{}_{\rm LS}$, we permit their absolute values to fluctuate within 0.5 to 2 times the reference scale ${\cal O}(M^{}_{\rm LS})$. Similarly, the elements $a$, $b$ and $c$ of the matrix $R$ are varied within the interval [0.5, 2]. The initial energy scale $\Lambda$ for RGE is set with the benchmark ratio $\Lambda/M^{}_1=100$. In practice, the dependence of the final results on this choice is weak, as the RGE effects depend on $\Lambda$ only logarithmically. To be consistent with the core idea of the LSS model (namely, $M^{}_{\rm LS} \ll M^{}_{\rm D} \ll M^{}_{\rm RS}$), we impose the constraint that $M^{}_{\rm LS}$ is at least two orders of magnitude smaller than $M^{}_{\rm D}$, and $M^{}_{\rm D}$ is in turn at least two orders of magnitude smaller than $M^{}_{\rm RS}$. The results indicate that the observed value of $Y^{}_{\rm B}$ can be successfully reproduced in both the NO and IO cases. For example, in the NO case with $M^{}_1 =1$ TeV, the observed value of $Y^{}_{\rm B}$ can be achieved when ${\cal O}(M^{}_{\rm LS})$ falls within the range $\sim 10^2$---$\sim 10^3$ eV, which corresponds to ${\cal O}(M^{}_{\rm D})$ values between $10^9$ and $10^8$ eV.

\subsection{Charged lepton flavor violation}

In the context of our model, the Yukawa interactions responsible for generating neutrino masses also give rise to charged lepton flavor violation (cLFV). Among various possible cLFV channels, the radiative decay $l_\alpha \to l_\beta \gamma $ is the most widely studied one, as it provides the most stringent experimental limits and is expected to exhibit the largest branching ratios. According to the MEG collaboration \cite{MEG:2016leq}, one has the experimental limit ${\rm BR}(\mu \to e \gamma) < 4.2 \times 10^{-13}$. Therefore, the evaluation of branching ratios for these decays within our framework not only acts as a key consistency check of the model but also establishes a direct connection between the neutrino mass generation mechanism and experimentally accessible low-energy signals.

Within our framework, ${\rm BR}(l_\alpha \to l_\beta \gamma)$ can be calculated according to \cite{Forero:2011pc}
\begin{eqnarray}
{\rm BR}(l_\alpha \to l_\beta \gamma)= \frac{\alpha^3_W s_W^2}{256 \pi^2}\frac{m_{l_\alpha}^5}{M_W^4}
\frac{1}{\Gamma_{l_{\alpha}}}|G_{\alpha \beta}^W|^2 \; ,
\label{3.12}
\end{eqnarray}
where
\begin{eqnarray}
&& G_{\alpha \beta}^W=\sum_{i=1}^9 K^*_{\alpha i} K_{\beta i} G_\gamma^W \left(\frac{M^2_{i}}{M_W^2}\right) \; ,
 \nonumber \\
&& G_\gamma^W(x)=\frac{1}{12(1-x)^4}(10-43x+78x^2-49x^3+18x^3\ln{x}+4x^4) \;.
\label{3.13}
\end{eqnarray}
Here $ \alpha _W \equiv g_W^2/(4\pi)$ denotes the weak fine-structure constant constant, $s_W^2= \sin^2 \theta _W$ with $\theta _W$ being the Weinberg angle, $M_W$ the $W$-boson mass, $m_{l_\alpha}$ and $\Gamma_{l_{\alpha}}$ the mass and total decay width of the decaying charged lepton $l_\alpha$.
Finally, one has the matrix $K=\left(K_{L},K_{H}\right)$ with
\begin{eqnarray}
&& K_L = \left(I-\frac{1}{2}(M^{}_{\rm D}, M^{}_{\rm LS})^*(M_{NS}^{\prime -1})^{*} M_{NS}^{\prime -1} (M^{}_{\rm D}, M^{}_{\rm LS})^T\right) U^{\prime *} \; ,
 \nonumber \\
&& K_H = (M^{}_{\rm D}, M^{}_{\rm LS})^* (M_{NS}^{\prime -1})^{*} V \;.
\label{3.14}
\end{eqnarray}

\begin{figure*}
\centering
\includegraphics[width=6.5in]{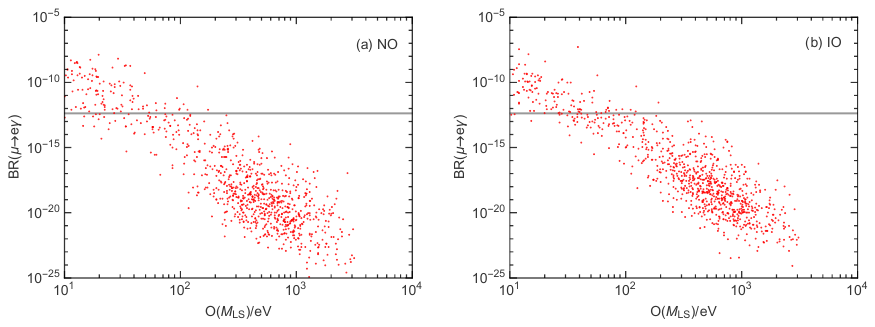}
\caption{ For the scenario studied in section~3.1, in the parameter space that allows for a reproduction of the observed value of $Y^{}_{\rm B}$, the allowed values of ${\rm BR}(\mu \to e \gamma)$ as functions of ${\cal O}(M^{}_{\rm LS})$ in the NO (a) and IO (b) cases. The horizontal line stands for the current upper bound on ${\rm BR}(\mu \to e \gamma)$. }
\label{fig2}
\end{figure*}

Now we investigate the consequences of the model with respect to the $\mu \to e \gamma $ process. In Figure~\ref{fig2}(a) and (b) (for the NO and IO cases, respectively), we have shown the allowed values of ${\rm BR}(\mu \to e \gamma)$ as functions of ${\cal O}(M^{}_{\rm LS})$. With the same parameter settings as in section~3.1, these results are obtained within the parameter space that allows for a successful reproduction of the observed baryon asymmetry of the Universe, with $M^{}_1$ varying in the range from  1 to 10 TeV. The results show that the parameter range of ${\cal O}(M^{}_{\rm LS}) \lesssim 300$ eV is  excluded by the constraint from the $\mu \to e \gamma $ process, leaving us with $300 \lesssim {\cal O}(M^{}_{\rm LS})/{\rm eV} \lesssim 3000$.

\section{Baryon/lepton asymmetry after electroweak symmetry breaking}

In this section, we study the scenario that the mass splitting for the two sterile neutrinos that form a Dirac pair naturally arise from the mass terms existing in $M^{}_{\rm D}$ and $M^{}_{\rm LS}$ which are generated after the EWSB. Given that there exists a narrow interval between the EWSB ($\simeq 159$ GeV) and sphaleron freeze-out ($\simeq 132$ GeV), according to the temperature where leptogenesis takes place, there are two distinct scenarios: in the first scenario, leptogenesis takes place between the EWSB and sphaleron freeze-out, in which case the generated lepton asymmetry can get converted into the baryon asymmetry via the sphaleron processes; in the second scenario, leptogenesis takes place after the sphaleron freeze-out, in which case the generated lepton asymmetry will not get converted into a baryon asymmetry.

\subsection{Leptogenesis before sphaleron freeze-out}

We first study the scenario that leptogenesis takes place between the EWSB and sphaleron freeze-out temperature, in which case the generated lepton asymmetry can get converted into the baryon asymmetry via the sphaleron processes.

Before the EWSB, there only exist the Yukawa interactions between active and sterile neutrinos (as well as the Higgs field) but not the mass terms relating them. When the Higgs field acquires the vacuum expectation value, these interactions produce the mass terms in $M_{\rm D}$ and $M_{\rm LS}$ (as mentioned below Eq.~(\ref{1.2})). These mass terms will make corrections to the sterile neutrino mass matrix after the seesaw diagonalization of the full neutrino mass matrix. To be specific, the effective mass matrix for the sterile neutrino sector is modified to \cite{SV}
\begin{eqnarray}
&&M^{\prime}_{\rm RS}  =  M^{}_{\rm RS} + \frac{1}{2} \left[ (M^{-1}_{\rm RS})^* (M^{}_{\rm D}, M^{}_{\rm LS})^\dagger (M^{}_{\rm D}, M^{}_{\rm LS})+ (M^{}_{\rm D}, M^{}_{\rm LS})^{T} (M^{}_{\rm D}, M^{}_{\rm LS})^* (M^{-1}_{\rm RS})^* \right] \;.
\label{4.1}
\end{eqnarray}
Like the RGE effects considered in section~3, these corrections will lead to non-zero mass terms $\mu_n$ and $\mu_s$ (as shown in Eq.~(\ref{3.3})):
\begin{eqnarray}
&&\mu^{}_n = \frac{1}{2} \left[(M^{-1}_{\rm RS})^*  (M^{\dagger}_{\rm LS} M^{}_{\rm D})+ (M^{\dagger}_{\rm LS} M^{}_{\rm D})^T (M^{-1}_{\rm RS})^\dagger \right] \;,
\nonumber\\
&&\mu^{}_s = \frac{1}{2} \left[(M^{-1}_{\rm RS})^\dagger  (M^{\dagger}_{\rm D} M^{}_{\rm LS})+ (M^{\dagger}_{\rm D} M^{}_{\rm LS})^T (M^{-1}_{\rm RS})^* \right] \;,
\label{4.2}
\end{eqnarray}
which just can break the exact mass degeneracy between the two sterile neutrinos that form a Dirac pair.
These terms are of the order ${\cal O}(M^{}_{\rm D} M^{}_{\rm LS}/M^{}_{\rm RS})$, which are just comparable to the light neutrino masses in magnitude. In spite of their smallness, $\mu_n$ and $\mu_s$ play a crucial role in enabling resonant enhancement of the CP asymmetry in the decays of sterile neutrinos.

\begin{figure*}
\centering
\includegraphics[width=6.5in]{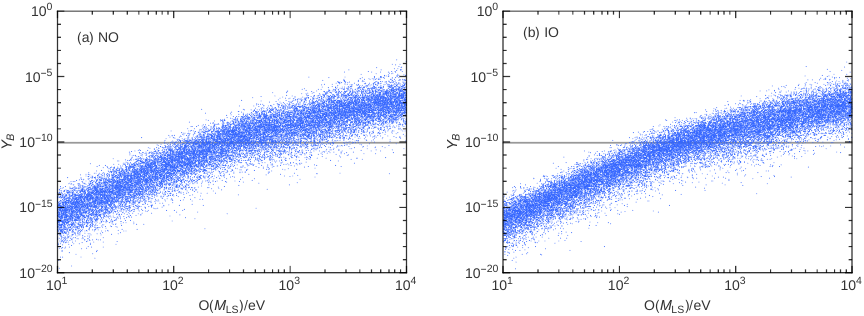}
\caption{ For the scenario studied in section~4.1 that only the EWSB contributes to the mass splittings for the sterile neutrinos, in the NO (a) and IO (b) cases, the allowed values of $Y^{}_{\rm B}$ as functions of ${\cal O}(M^{}_{\rm LS})$ for the benchmark value of $M^{}_1=150$ GeV. The horizontal line stands for the observed value of $Y^{}_{\rm B}$. }
\label{fig3}
\end{figure*}

\begin{figure*}
\centering
\includegraphics[width=6.5in]{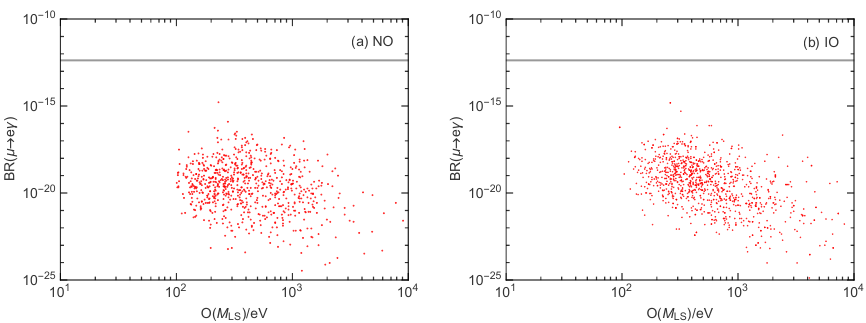}
\caption{ For the scenario studied in section~4.1 that only the EWSB contributes to the mass splittings for the sterile neutrinos, in the parameter space that allows for a reproduction of the observed value of $Y^{}_{\rm B}$, the allowed values of ${\rm BR}(\mu \to e \gamma)$ as functions of ${\cal O}(M^{}_{\rm LS})$ in the NO (a) and IO (b) cases. The horizontal line stands for the current experimental limit on ${\rm BR}(\mu \to e \gamma)$. }
\label{fig4}
\end{figure*}

Although the temperature window between the EWSB and sphaleron freeze-out is narrow, leptogenesis occurring within this interval can still efficiently generate the desired baryon asymmetry. To get the resulting values of $Y^{}_{\rm B}$, we solve the relevant Boltzmann equations by employing the ULYSSES package \cite{Granelli:2020pim} and follow the dynamical evolution down to the sphaleron freeze-out temperature. For the present scenario, Figure~\ref{fig3}(a) and (b) (for the NO and IO cases, respectively) have shown the allowed values of $Y^{}_{\rm B}$ as functions of ${\cal O}(M^{}_{\rm LS})$ for the benchmark value of $M^{}_1 = 150$ GeV. These results are obtained for the same parameter settings as in section~3.1. The results indicate that the observed value of $Y^{}_{\rm B}$ can be achieved when ${\cal O}(M^{}_{\rm LS})$ falls within the range $\sim 10^2$---$ \sim 10^4$ eV in both the NO and IO cases. For this scenario, we have also investigated the consequences of the model with respect to the $\mu \to e \gamma $ process. The results are shown in Figure~\ref{fig4}(a) and (b) (for the NO and IO cases, respectively). It turns out that, for the parameter space consistent with the observed value of $Y^{}_{\rm B}$, the allowed values of ${\rm BR}(\mu \to e \gamma)$ are well below the current upper bound.

\begin{figure*}
\centering
\includegraphics[width=6.5in]{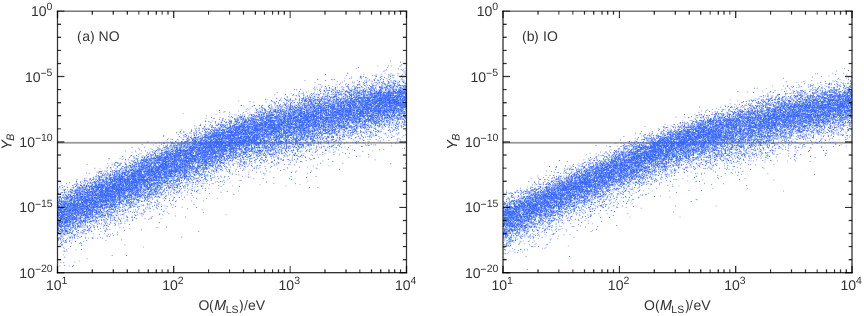}
\caption{ For the scenario studied in section~4.1 that both the RGE effects and the EWSB contribute to the mass splittings for the sterile neutrinos, the allowed values of $Y^{}_{\rm B}$ as functions of ${\cal O}(M^{}_{\rm LS})$ in the NO (a) and IO (b) cases. The horizontal line stands for the observed value of $Y^{}_{\rm B}$. }
\label{fig5}
\end{figure*}

In fact, the contributions of the RGE effects and the EWSB to the mass splittings for the sterile neutrinos may coexist. The contributions of the RGE effects to their mass splittings are in the order of ${\cal O}\left(M^{}_{\rm D} M^{}_{\rm LS}M^{}_{\rm RS}\ln \left(\Lambda/M^{}_1\right)/v^2\right)$, which are also comparable to the light neutrino masses when the sterile neutrinos are around the electroweak scale. But as the sterile neutrino masses decrease, these contributions become subdominant compared to those from the EWSB.
For completeness, taking into account both the contributions of the RGE effects and the EWSB to the mass splittings for the sterile neutrinos, we have shown the allowed values of $Y^{}_{\rm B}$ as functions of ${\cal O}(M^{}_{\rm LS})$ in Figure~\ref{fig5}(a) and (b) (for the NO and IO cases, respectively). It is found that the parameter space that allows for a reproduction of the observed value of $Y^{}_{\rm B}$ is very similar to that shown in Figure~\ref{fig3}, where only the contributions of the EWSB to the mass splittings for the sterile neutrinos were considered. The reason is that although the combined EWSB and RGE effects increase the total mass splittings for the sterile neutirnos and thereby enhance the CP asymmetries for their decay processes, they also strengthen the washout effects which have the opposite effect, as described by Eq.~(\ref{2.4}).

\subsection{Leptogenesis after sphaleron freeze-out}

 Then, we study the scenario that leptogenesis takes place after the sphaleron freeze-out, in which case the generated lepton asymmetry will not get converted into a baryon asymmetry. In this connection, we explore the possibility that a large lepton asymmetry may arise (while a baryon asymmetry of comparable magnitude will not accompany).

Our motivation for considering such an intriguing possibility is partially inspired by recent cosmological observations that indicate the possible existence of a large lepton asymmetry in the early Universe: recently, the EMPRESS collaboration reported a new determination of the primordial ${}^4\mathrm{He}$ abundance \cite{Matsumoto:2022tlr}
\begin{eqnarray}
Y_p = 0.2370^{+0.0034}_{-0.0033} \;,
\label{1.9}
\end{eqnarray}
which is below the prediction of standard Big Bang nucleosynthesis by about the $3\sigma$ level. A natural explanation for this discrepancy invokes a lepton asymmetry $Y^{}_{\rm L}\sim 5 \times 10^{-3}$, which is much larger than the observed baryon asymmetry of the Universe \cite{BTV, EIM}. While such an indication is only suggestive at the present stage and it may disappear with future observations, it is still tantalising to explore the possibility that there exists a large primordial lepton asymmetry for the following three considerations (for a review see, e.g., Ref.~\cite{LM}): 1) although a lepton asymmetry may leave observable imprints on the evolution history of the Universe, and thus can be constrained by cosmological observations for Big Bang Nucleosynthesis, Cosmic Microwave Background and Large-Scale Structure (for recent studies see, e.g., Refs.~\cite{FP, LY}), current constraints on it are weaker than the baryonic measurement by many orders of magnitude, so it is in principle allowed to be much larger than its baryonic counterpart. 2) A large lepton asymmetry has the potential to alleviate some cosmological tensions (e.g., the infamous Hubble tension \cite{hubble}) \cite{Barenboim:2016lxv}-\cite{LY2}. 3) A large lepton asymmetry can also facilitate the resonant production of sterile-neutrino dark matter via the Shi-Fuller mechanism \cite{Shi:1998km}.

Intuitively, it is natural to expect that lepton and baryon asymmetries would be comparable in size, given the role of sphaleron transitions in the early Universe (see Eq.~(\ref{2.1})). In fact, various theoretical frameworks have been developed where the final lepton asymmetry could be significantly larger than its baryonic counterpart. In such scenarios, a large lepton asymmetry is typically produced at temperatures below the sphaleron freeze-out (i.e., $T \simeq 132$ GeV \cite{sphaleron}), via resonant leptogenesis \cite{BD}, freeze-in leptogenesis \cite{AS, ABS}, the Affleck-Dine mechanism \cite{CCG}-\cite{Dine}, the decay of topological defects \cite{BRS}, Q-ball decays \cite{KTY, KM}, or decays of dark matter \cite{BDS}. Interestingly, in Refs.~\cite{CEKL}-\cite{CEKL2}, by introducing temperature-dependent masses for sterile neutrinos, it becomes possible to realize two distinct phases of leptogenesis occurring at separate temperature scales. One phase takes place prior to sphaleron freeze-out, enabling it to generate the observed baryon asymmetry of the Universe; while the other occurs after sphaleron freeze-out, thereby allowing for the production of a large lepton asymmetry.
Furthermore, there are scenarios where significant flavor-specific lepton asymmetries are generated prior to sphaleron freeze-out, although the total lepton asymmetry is effectively zero \cite{MMR}-\cite{ZZH2}.

In the following, by making use of the observation that in the LSS model the EWSB automatically induces the desired mass splitting for the two sterile neutrinos that form a Dirac pair (so that a large lepton asymmetry may naturally arise from the resonant leptogenesis scenario) and the sphaleron processes freeze out almost at the same time (so that a baryon asymmetry of comparable magnitude will probably not accompany), we give a tentative explanation for the possible existence of a large lepton asymmetry. As discussed above, since the RGE effect is subdominant below the electroweak scale, we will not include the RGE effects in this scenario.

\begin{figure*}
\centering
\includegraphics[width=6.5in]{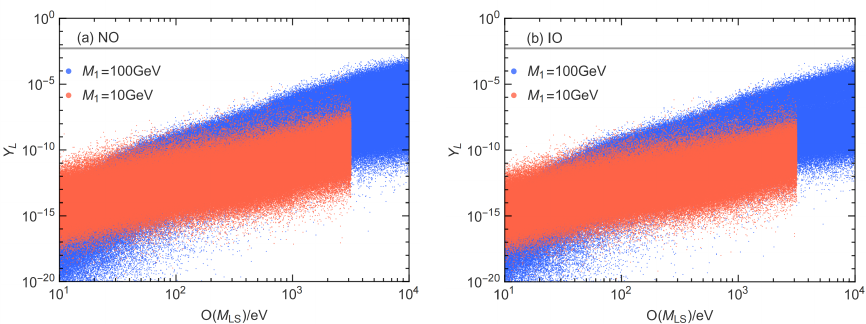}
\caption{ For the scenario studied in section~4.2, in the NO (a) and IO (b) cases, the allowed values of $Y^{}_{\rm L}$ as functions of ${\cal O}(M^{}_{\rm LS})$ for some benchmark values of $M^{}_1$. The horizontal line stands for the value of $Y^{}_{\rm L}$ indicated by the EMPRESS results. }
\label{fig6}
\end{figure*}

For the present scenario, Figure~\ref{fig6}(a) and (b) (for the NO and IO cases, respectively) have shown the allowed values of $Y^{}_{\rm L}$ as functions of ${\cal O}(M^{}_{\rm LS})$ for the benchmark values of $M^{}_1 = 10$ and $100$ GeV. These results are obtained for the same parameter settings as in section~3.1. The results show that in both the NO and IO cases, a large lepton asymmetry can be generated. In particular, for the NO case with $M^{}_1 \sim 100$ GeV, one obtains a maximum achievable lepton asymmetry of $Y_{\rm L}^{\rm max} \sim 1.8 \times 10^{-3}$. In the IO case with $M^{}_1 \sim 100$ GeV, the maximum achievable lepton asymmetry turns out to be $Y_{\rm L}^{\rm max}\sim 1.3 \times 10^{-3}$. Although the maximum lepton asymmetry achievable in this scenario is somewhat smaller than the value suggested by the EMPRESS results (which remains indicative at this stage), the framework nonetheless offers a natural mechanism for producing a substantial lepton asymmetry that is consistent with the observed baryon asymmetry. Moreover, if the requirement of a strong hierarchy between $M^{}_{\rm LS}$ and $M^{}_{\rm D}$ is relaxed, the resulting lepton asymmetry can be further enhanced, in particular when they are of comparable scales.

\section{Summary}

The type-I seesaw model provides an elegant framework that simultaneously explains the smallness of neutrino masses and the origin of the baryon asymmetry of the Universe via the leptogenesis mechanism. However, in order to naturally generate the small neutrino masses and observed baryon asymmetry, the right-handed neutrinos it introduces are typically required to be extremely heavy ($\gtrsim 10^9$ GeV), placing them far beyond the reach of any foreseeable experimental detection. In this connection, the linear seesaw model offers a promising alternative, which introduces additional singlet fermions $S^{}_{I}$. It can naturally accommodate relatively low  (TeV or so) sterile neutrino masses (while the smallness of neutrino masses can be attributed to a slight breaking of lepton number conservation), thereby enhancing their prospects for experimental accessibility. However, in generic LSS models, the sterile neutrinos typically arrange themselves in Dirac pairs with exact mass degeneracies, which precludes the generation of CP asymmetries from their interplay. But such a feature of the LSS model can serve as a unique starting point for the realization of resonant leptogenesis scenarios: if the sterile neutrino masses receive certain corrections so that their degeneracy is lifted to an appropriate extent, then the resonant leptogenesis scenario will be naturally realized, thereby enabling successful leptogenesis at low energy scales.

In this paper, we have explored two well-motivated approaches to generate the desired mass splitting for the two sterile neutrinos that form a Dirac pair: the renormalization group evolution effect and the electroweak symmetry breaking. These two approaches have the common merits that they are spontaneous and do not need to introduce additional parameters. However, when the latter approach takes effect the sphaleron processes will freeze out almost at the same time. This means that the lepton asymmetry generated from the latter approach probably will not get converted into a baryon asymmetry of the same order of magnitude. We have just made use of such a feature of the LSS model to give a tentative explanation for the possible existence of a large lepton asymmetry as indicated by the recent result from the EMPRESS collaboration, without contradicting the observed baryon asymmetry.

In the first scenario, we have taken advantage of the RGE effects to generate the desired mass splitting for the two sterile neutrinos that form a Dirac pair, naturally realizing the resonant leptogenesis. Figure~\ref{fig1} (a) and (b) demonstrate that in both the NO and IO cases, the observed value of $Y^{}_{\rm B}$ can be successfully reproduced. For example, in the NO case with $M^{}_1 =1$ TeV, the observed value of $Y^{}_{\rm B}$ can be achieved when ${\cal O}(M^{}_{\rm LS})$ falls within the range $\sim 10^2$---$\sim 10^3$ eV.
For this scenario, we have also investigated the consequences of the model with respect to the $\mu \to e \gamma $ process. The results show that the parameter range of ${\cal O}(M^{}_{\rm LS}) \lesssim 300$ eV is excluded by the constraint from the $\mu \to e \gamma $ process, leaving us with $300 \lesssim {\cal O}(M^{}_{\rm LS})/{\rm eV} \lesssim 3000$.

In the second scenario, the mass terms arising from the EWSB induce tiny mass splittings within each pair of pseudo-Dirac sterile neutrinos, thereby enabling resonant leptogenesis after the EWSB. If  leptogenesis happens to take place within the temperature window between the EWSB ($\simeq 159$ GeV) and sphaleron freeze-out ($\simeq 132$ GeV), the generated lepton asymmetry can still get efficiently converted into a baryon asymmetry. As shown in Figure~\ref{fig3}(a) and (b), the observed value of $Y^{}_{\rm B}$ can be successfully reproduced in both the NO and IO cases for ${\cal O}(M^{}_{\rm LS})$ in the range $\sim 10^2$---$ \sim 10^4$ eV. On the other hand, if leptogenesis takes place after the sphaleron freeze-out, the generated lepton asymmetry will not get converted into a baryon asymmetry. By making use of this fact, we have explored the possibility that a large lepton asymmetry may arise (while a baryon asymmetry of comparable magnitude will not accompany). Figure~\ref{fig6} (a) and (b) show that for $M^{}_1 \sim 100$ GeV in the NO (IO) case, one obtains a maximum achievable lepton asymmetry of $Y_{\rm L}^{\rm max} \sim 1.8 \times 10^{-3}$ ($1.3 \times 10^{-3}$). Although these results are somewhat smaller than the value suggested by the EMPRESS results (which remains indicative at this stage), the framework nonetheless offers a natural mechanism for producing a substantial lepton asymmetry that is consistent with the observed baryon asymmetry.

\vspace{0.5cm}

\underline{Acknowledgments} \vspace{0.2cm}

This paper is dedicated to Zhao's beloved mother, who passed away suddenly during the writing of this paper. This work was supported in part by the National Natural Science Foundation of China under Grant No.~12475112, Liaoning Revitalization Talents Program under Grant No.~XLYC2403152, and the Basic Research Business Fees for Universities in Liaoning Province under Grant No.~LJ212410165050.


\begin{thebibliography}{99}


\bibitem{Planck:2018vyg} N.~Aghanim \textit{et al.} [Planck], Astron. Astrophys. \textbf{641}, A6 (2020).

\bibitem{leptogenesis} M. Fukugita and T. Yanagida, Phys. Lett. B {\bf 174}, 45 (1986).

\bibitem{Lreview1} W. Buchmuller, R. D. Peccei and T. Yanagida, Ann. Rev. Nucl. Part. Sci. {\bf 55}, 311 (2005).

\bibitem{Lreview2} W. Buchmuller, P. Di Bari and M. Plumacher, Annals Phys. {\bf 315}, 305 (2005).

\bibitem{Lreview3} S. Davidson, E. Nardi and Y. Nir, Phys. Rept. {\bf 466}, 105 (2008).

\bibitem{Lreview4} D. Bodeker and W. Buchmuller, Rev. Mod. Phys. {\bf 93}, 035004 (2021).

\bibitem{seesaw1} P. Minkowski, Phys. Lett. B {\bf 67}, 421 (1977).

\bibitem{seesaw2} M. Gell-Mann, P. Ramond and R. Slansky, in Supergravity, edited by P. van Nieuwenhuizen and D. Freedman, (North-Holland, 1979), p. 315.

\bibitem{seesaw3}  T. Yanagida, in Proceedings of the Workshop on the Unified Theory and the Baryon Number in the Universe, edited by O. Sawada and A. Sugamoto (KEK Report No. 79-18, Tsukuba, 1979), p. 95.

\bibitem{seesaw4} R. N. Mohapatra and G. Senjanovic, Phys. Rev. Lett. {\bf 44}, 912 (1980).

\bibitem{seesaw5} J. Schechter and J. W. F. Valle, Phys. Rev. D {\bf22}, 2227 (1980).

\bibitem{sphaleron} M. D'Onofrio, K. Rummukainen and A. Tranberg, Phys. Rev. Lett. {\bf 113}, 141602 (2014).

\bibitem{DI} S. Davidson and A. Ibarra, Phys. Lett. B {\bf 535}, 25 (2002).

\bibitem{resonant1} A. Pilaftsis, Phys. Rev. D {\bf 56}, 5431 (1997).

\bibitem{resonant2} A. Pilaftsis and T. E. J. Underwood, Nucl. Phys. B {\bf 692}, 303 (2004).

\bibitem{LSS1} E.~K.~Akhmedov, M.~Lindner, E.~Schnapka and J.~W.~F.~Valle, Phys. Lett. B {\bf 368}, 270 (1996).

\bibitem{LSS2} E.~K.~Akhmedov, M.~Lindner, E.~Schnapka and J.~W.~F.~Valle, Phys. Rev. D {\bf 53},   2752 (1996).

\bibitem{LSS3} M.~Malinsky, J.~C.~Romao and J.~W.~F.~Valle, Phys. Rev. Lett. \textbf{95}, 161801 (2005).

\bibitem{FMN} E.~Fernandez-Martinez, X.~Marcano and D.~Naredo-Tuero, JHEP {\bf 03}, 057 (2023).


\bibitem{pseudo1} L. Wolfenstein, Nucl. Phys. B {\bf 186}, 147 (1981).

\bibitem{pseudo2} S. T. Petcov, Phys. Lett. B {\bf 110}, 245 (1982).

\bibitem{pseudo3} J. W. F. Valle and M. Singer, Phys. Rev. D {\bf 28}, 540 (1983).

\bibitem{pseudo4} M. Kobayashi and C. S. Lim, Phys. Rev. D {\bf 64}, 013003 (2001).


\bibitem{flavor1} A. Abada, S. Davidson, F. X. Josse-Michaux, M. Losada and A. Riotto, JCAP {\bf0604}, 004 (2006).

\bibitem{flavor2} E. Nardi, Y. Nir, E. Roulet and J. Racker, JHEP {\bf0601}, 164 (2006).

\bibitem{Yv} S.~Blanchet, T.~Hambye and F.~X.~Josse-Michaux, JHEP {\bf 04}, 023 (2010).

\bibitem{zB} W.~Buchmuller, P.~Di Bari and M.~Plumacher, Annals Phys. {\bf 315}, 305 (2005).

\bibitem{ADEFHV} K.~Agashe, P.~Z.~Du, M.~Ekhterachian, C.~S.~Fong, S.~Hong and L.~Vecchi, JHEP {\bf 04}, 029 (2019).

\bibitem{Brdar:2015jwo} V.~Brdar, M.~K{\"o}nig and J.~Kopp, Phys. Rev. D \textbf{93}, 093010 (2016).


\bibitem{global1} I.~Esteban, M.~C.~Gonzalez-Garcia, M.~Maltoni, I.~Martinez-Soler, J.~P.~Pinheiro and T.~Schwetz,JHEP \textbf{12}, 216 (2024).

\bibitem{global2} F.~Capozzi, W.~Giar{\`e}, E.~Lisi, A.~Marrone, A.~Melchiorri and A.~Palazzo, Phys. Rev. D \textbf{111} , 093006 (2025).

\bibitem{CI} J. A. Casas and A. Ibarra, Nucl. Phys. B {\bf 618}, 171 (2001).

\bibitem{Forero:2011pc} D.~V.~Forero, S.~Morisi, M.~Tortola and J.~W.~F.~Valle, JHEP \textbf{09}, 142 (2011).

\bibitem{xing} H.~C.~Han and Z.~Z.~Xing, Nucl. Phys. B {\bf 973}, 115609 (2021).

\bibitem{xingPR} Z.~Z.~Xing, Phys. Rep. {\bf 854}, 1 (2020).

\bibitem{MEG:2016leq} A.~M.~Baldini \textit{et al.} [MEG], Eur. Phys. J. C \textbf{76}, 434 (2016).

\bibitem{SV} J. Schechter and J. W. F. Valle, Phys. Rev. D {\bf25}, 774 (1982).

\bibitem{Granelli:2020pim} A.~Granelli, K.~Moffat, Y.~F.~Perez-Gonzalez, H.~Schulz and J.~Turner, Comput. Phys. Commun. \textbf{262}, 107813 (2021).

\bibitem{Matsumoto:2022tlr} A.~Matsumoto, M.~Ouchi, K.~Nakajima, M.~Kawasaki, K.~Murai, K.~Motohara, Y.~Harikane, Y.~Ono, K.~Kushibiki and S.~Koyama, Astrophys. J. {\bf941}, 167 (2022).

\bibitem{BTV} A.~K.~Burns, T.~M.~P.~Tait and M.~Valli, Phys. Rev. Lett. {\bf 130}, 131001 (2023).

\bibitem{EIM} M.~Escudero, A.~Ibarra and V.~Maura, Phys. Rev. D {\bf 107}, 035024 (2023).

\bibitem{LM} M.~Lattanzi and M.~Moretti, Symmetry {\bf 16}, 1657 (2024).

\bibitem{FP} J.~Froustey and C.~Pitrou, Phys. Rev. D {\bf 110}, 103551 (2024).

\bibitem{LY} Y.~Z.~Li and J.~H.~Yu, JHEP {\bf 06}, 213 (2025).

\bibitem{hubble} E.~Di~Valentino {\it et al.}, Class. Quant. Grav. {\bf 38}, 153001 (2021).

\bibitem{Barenboim:2016lxv} G.~Barenboim, W.~H.~Kinney and W.~I.~Park, Eur. Phys. J. C \textbf{77}, 590 (2017).

\bibitem{Yeung:2020zde} S.~Yeung, K.~Lau and M.~C.~Chu, JCAP \textbf{04}, 024 (2021).

\bibitem{Seto:2021tad} O.~Seto and Y.~Toda, Phys. Rev. D \textbf{104},  063019 (2021).

\bibitem{Kumar:2022vee} S.~Kumar, R.~C.~Nunes and P.~Yadav, JCAP \textbf{09}, 060 (2022).

\bibitem{LY2}  Y.~Z.~Li and J.~H.~Yu, arXiv:2501.13153.

\bibitem{Shi:1998km} X.~D.~Shi and G.~M.~Fuller, Phys. Rev. Lett. \textbf{82}, 2832 (1999).

\bibitem{BD} D.~Borah and A.~Dasgupta, Phys. Rev. D {\bf 108}, 035015 (2023).

\bibitem{AS} T.~Asaka and M.~Shaposhnikov, Phys. Lett. B {\bf 620}, 17 (2005).

\bibitem{ABS} T.~Asaka, S.~Blanchet and M.~Shaposhnikov, Phys. Lett. B {\bf 631}, 151 (2005).

\bibitem{CCG} A.~Casas, W.~Y.~Cheng and G.~Gelmini, Nucl. Phys. B {\bf 538}, 297 (1999).

\bibitem{DK} A.~D.~Dolgov and D.~P.~Kirilova, Sov. J. Nucl. Phys. {\bf 51}, 172 (1990).

\bibitem{Affleck:1984fy} I.~Affleck and M.~Dine, Nucl. Phys. B \textbf{249}, 361-380 (1985).

\bibitem{Dine} M.~Dine, L.~Randall and S.~D.~Thomas, Nucl. Phys. B \textbf{458}, 291 (1996).

\bibitem{BRS} B.~Bajc, A.~Riotto and G.~Senjanovic, Phys. Rev. Lett. {\bf 81}, 1355 (1998).

\bibitem{KTY} M.~Kawasaki, F.~Takahashi and M.~Yamaguchi, Phys. Rev. D {\bf 66}, 043516 (2002).

\bibitem{KM} M.~Kawasaki and K.~Murai, JCAP \textbf{08}, 041 (2022).

\bibitem{BDS} D.~Borah, N.~Das and I.~Saha, arXiv:2410.00096.

\bibitem{CEKL} Y.~ChoeJo, K.~Enomoto, Y.~Kim and H.~S.~Lee, JHEP {\bf 03}, 003 (2024).

\bibitem{BDS2} D.~Bhandari, A.~Datta and A.~Sil, Phys. Rev. D {\bf 110}, 115008 (2024).

\bibitem{CEKL2} Y.~ChoeJo, K.~Enomoto, Y.~Kim and H.~S.~Lee, Phys. Rev. D {\bf 111}, 055026 (2025).

\bibitem{MMR} J. March-Russell, H. Murayama and A. Riotto, JHEP {\bf 11}, 015 (1999).

\bibitem{MSY} K. Mukaida, K. Schmitz and M. Yamada, Phys. Rev. Lett. {\bf 129}, 011803 (2022).

\bibitem{ZZH1} Z.~H.~Zhao, J.~Zhang and X.~Y.~Wu, JHEP {\bf 09}, 094 (2024).

\bibitem{ZZH2} Y.~Shao and  Z.~H.~Zhao, Phys. Rev. D {\bf 111}, 035011 (2025).




















\end{thebibliography}

\end{document}